\documentclass[%
aps,
11pt,
floatfix,
reprint,
onecolumn,
tightenlines,
superscriptaddress,
eqsecnum,
preprintnumbers,
physrev,
]{revtex4-2}

\usepackage{isomath}
\usepackage{amsmath,amsthm}
\usepackage{amsbsy}
\usepackage{amssymb}
\usepackage{amscd}
\usepackage{amsfonts}
\usepackage{stmaryrd}
\usepackage{siunitx}
\usepackage{euscript}
\usepackage[utf8]{inputenc}
\usepackage[T1]{fontenc}
\usepackage{newtxtext} 
\everymath{\displaystyle}
\usepackage{exscale}

\usepackage{graphicx}
\usepackage{boxedminipage}
\usepackage{calc}
\usepackage[dvipsnames]{xcolor}
\graphicspath{ {media/} }
\usepackage[caption=false,justification=centerlast]{subfig}

\usepackage{setspace}
\usepackage{enumitem}
\setitemize{noitemsep,topsep=0pt,parsep=0pt,partopsep=0pt}
\setenumerate{noitemsep,topsep=0pt,parsep=0pt,partopsep=0pt}
\setdescription{noitemsep,topsep=0pt,parsep=0pt,partopsep=0pt}

\usepackage[colorinlistoftodos, color=green!40,prependcaption]{todonotes}

\usepackage{soul} % highlighting

\usepackage{hyperref}
\usepackage[normalem]{ulem}

\usepackage[small]{titlesec}

\titlespacing*{\section}{0pt}{12pt plus 4pt minus 2pt}{2pt plus 2pt minus 2pt}
\titlespacing*{\subsection}{0pt}{12pt plus 4pt minus 2pt}{2pt plus 2pt minus 2pt}
\titlespacing*\subsubsection{0pt}{12pt plus 4pt minus 2pt}{2pt plus 2pt minus 2pt}
\titlespacing*\paragraph{0pt}{12pt plus 4pt minus 2pt}{2pt plus 2pt minus 2pt}

\makeatletter

    \renewcommand*{\p@subsection}{}
    
    \renewcommand*{\p@subsubsection}{}
\makeatother

\numberwithin{figure}{section}

\usepackage{cleveref}
\crefname{figure}{Fig.}{Fig.}
\Crefname{figure}{Fig.}{Fig.}
\crefname{section}{Section}{Sections}
\crefname{table}{Table}{Tables}
\crefname{equation}{Eq.}{Eqs.}

\setuptodonotes{inline}
\usepackage{isomath}
\usepackage{amsmath}
\usepackage{amssymb}
\usepackage{amscd}
\usepackage{amsfonts}

\begin{document}

%%%%%%%%%%%%%%%%%%%%%
%%%%%%%%%%%%%%%%%%%%%
%%%%%%%%%%%%%%%%%%%%%
%%%%%%%%%%%%%%%%%%%%%

\preprint{To appear in Journal of Applied Mechanics (\href{https://doi.org/10.1115/1.4068059}{DOI:10.1115/1.4068059})}

\title{The Roles of Size, Packing, and Cohesion in the Emergence of Force Chains in Granular Packings}

\author{Ankit Shrivastava}
    \email{shrivastavaa@ornl.gov}
    \affiliation{Oak Ridge National Laboratory, Oak Ridge, Tennessee}

\author{Kaushik Dayal}
    \affiliation{Department of Civil and Environmental Engineering, Carnegie Mellon University}
    \affiliation{Center for Nonlinear Analysis, Department of Mathematical Sciences, Carnegie Mellon University}
    \affiliation{Department of Mechanical Engineering, Carnegie Mellon University}
        
\author{Hae Young Noh}
    \affiliation{Department of Civil and Environmental Engineering, Stanford University}

\date{\today}

%%%%%%%%%%%%%%%%%%%%%
%%%%%%%%%%%%%%%%%%%%%
%%%%%%%%%%%%%%%%%%%%%
%%%%%%%%%%%%%%%%%%%%%

\begin{abstract}	
This study investigates computationally the impact of particle size disparity and cohesion on force chain formation in granular media. 
The granular media considered in this study are bi-disperse systems under uniaxial compression, consisting of spherical, frictionless particles that interact through a modified Hookean model.
Force chains in granular media are characterized as networks of particles that meet specific criteria for particle stress and inter-particle forces.
The computational setup decouples the effects of particle packing on force chain formations, ensuring an independent assessment of particle size distribution and cohesion on force chain formation.
The decoupling is achieved by characterizing particle packing through the radial density function, which enables the identification of systems with both regular and irregular packing.
The fraction of particles in the force chains network is used to quantify the presence of the force chains.

The findings show that particle size disparity promotes force chain formation in granular media with nearly-regular packing (i.e., an almost-ordered system).
However, as particle size disparities grow, it promotes irregular packing (i.e., a disordered systems), leading to fewer force chains carrying larger loads than in ordered systems.
Further, it is observed that the increased cohesion in granular systems leads to fewer force chains irrespective of particle size or packing.
\end{abstract}

\maketitle

%%%%%%%%%%%%%%%%%%%%%
%%%%%%%%%%%%%%%%%%%%%
%%%%%%%%%%%%%%%%%%%%%
%%%%%%%%%%%%%%%%%%%%%

\section{Introduction}\label{sec:intro}

Granular media are a class of heterogeneous materials that play an essential role in stabilizing natural and industrial structures, such as river banks and building foundations.
It consists of discrete non-Brownian particles (grains) such as powders, sand, and glass beads that interact primarily through contact. 
It is well established that variations in the micro-scale properties of heterogeneous materials can result in localized regions of high stress under external loads. 
Accurately predicting these localized stresses and their impact on the macro-scale behavior of materials remains a key challenge.
In granular media, micro-scale variations occur when particle properties such as size, shape, inter-particle interactions, or packing are heterogeneous. 
The spatial variation in properties leads to localized stress patterns known as force chains that have been effectively visualized using various photoelastic experiments \cite{wang2020connecting, schneider1978molecular, radjai1996force, radjai1998bimodal, zhang2014force}.

In existing research, the force chains in granular media are defined as clusters of granular particles that transmit significantly higher loads than their neighboring particles.
The particles in force chains create percolating networks that extend over distances spanning several particle diameters, effectively serving as a structural backbone for the granular media.
Extensive research has investigated the distribution of inter-particle forces and transmission of external loads in force chains. 
These studies reveal a strong correlation between the characteristics of force chains and the bulk properties of the granular media.
For example, it was observed that the stability of the force chains is closely related to the load-bearing capacity \cite{li2016load, zhang2017role, tordesillas2009modeling, tordesillas2007force} and soil liquefaction \cite{wang2017liquefaction}.
As discussed in the works \citep{buarque2021unearthing, pal2021tunnel}, force chains can also reveal phenomena in granular media that micro- and continuum-scale models do not fully explain.
Therefore, studying the cause of force chain formation is crucial for understanding the behavior of granular media under various conditions and preventing failures in large-scale structures, such as buildings and their foundations.

Particle properties play a crucial role in shaping the characteristics of force chains within granular media. 
Numerous studies have explored the influence of loading conditions and granular properties on force chain characteristics—such as force distribution and network structure \cite{radjai1996force, majmudar2005contact, chen2020quantitative, antony2000evolution, an2013force, kondic2012topology, dijksman2018characterizing}.
For example, research by \cite{sun2010understanding} and \cite{kondic2009topology} have shown that particle packing fraction and friction play diverse roles in shaping force chain networks. 
Work by \cite{binaree2019effects} and \cite{azema2012force} demonstrated that granular particle shape significantly impacts the distribution of inter-particle forces, eventually affecting the force chains network.
Similarly, it was observed by \cite{buarque2021unearthing} that cohesion plays a vital role in the stability of granular media during excavation by redistributing inter-particle forces throughout the process.

Existing research has primarily examined the influence of particle properties on force chains in granular systems with random packing; however, packing can significantly influence force chain characteristics.
For example, in a crystalline packing of identical frictionless granular particles, force chains do not form under uniaxial compression because the packing allows forces to be transmitted evenly.
However, force chains can still form in a granular system with identical frictionless, cohesion-less particles but with arbitrary packing.
Hence, conclusions regarding the connection between the micro-scale properties and the characteristics of force chains will be conditioned on disorder in granular packing.
We refer to disorder in a granular system as any deviation from a crystalline packing (ordered system) caused by disparities in particle size, shape, inter-particle interactions, or other factors that lead to irregular packing.
In this work, we refer to the significant deviation of packing from an ordered system as a positional disorder and the minor deviation in packing from an ordered system caused by granular size disparity as a size disorder.
The particle packing is nearly crystalline in a size-disordered system, whereas the packing is entirely random in a positional-disordered system \cite{rosenbaum2019effects,rosenbaum2019surfactant}.

This work discusses the role of particle size variations and cohesion in forming force chains in a bi-disperse granular medium while decoupling the influence of packing on the force chains. 
This aim is to observe the transition in force chain formations in granular systems as the micro-scale properties are changed.
Size variation and cohesion in granular systems are quantified by two scalar variables: disorder and cohesion factors.
Since size variation introduces disorder in the system, we refer to the corresponding scalar variable as the disorder factor. 
Additionally, we classify the types of disorder in the system into size disorder and positional disorder to better understand whether size variation or particle assembly plays a more significant role.
Finally, we independently investigated the effects of size variation and cohesion on force chains by varying one property while holding the other constant.
While the general observations in this study regarding the effects of cohesion and disorder are consistent with existing literature, the proposed computational setup and characterization enable a quantitative assessment of their influence on force chains.

\paragraph*{Organization.}

\cref{sec:method} presents the contact model for simulating particle interactions, characterizing disorders, cohesion, and force chains in granular media.
\cref{sec:Ex_setup} discusses the computational setup for the granular media with various disorders and cohesion.
\cref{sec:results} discusses the observations regarding the impact of disorder and cohesion on the force chain characteristics.

\section{Computational Methodology}\label{sec:method}

The effects of disorder and cohesion on force chains are studied by simulating various two-dimensional granular systems with varying levels of disorder and cohesion.
The system comprises spherical, rigid, frictionless particles that interact according to a contact model outlined in  \cref{sec:contactmodel}.
The particles in the systems are categorized into two different types based on their respective sizes.
The difference in particle sizes quantifies the disorder, while cohesion is represented by a single scalar value, both described in  \cref{sec:characterize}.
The inter-particle forces and stresses are computed using the contact model under uniaxial compression, as detailed in  \cref{sec:simulate_setup}.
Finally, once the system reaches equilibrium, the inter-particle force values are used to characterize force chains, as explained in \cref{sec:forcechain_criteria}.

\subsection{Contact Model}\label{sec:contactmodel}

\begin{figure*}
	\centering
        \subfloat[Illustrative diagram of the modified Hookean model. \label{fig:illustrations}
	]{\includegraphics[width=.4\textwidth]{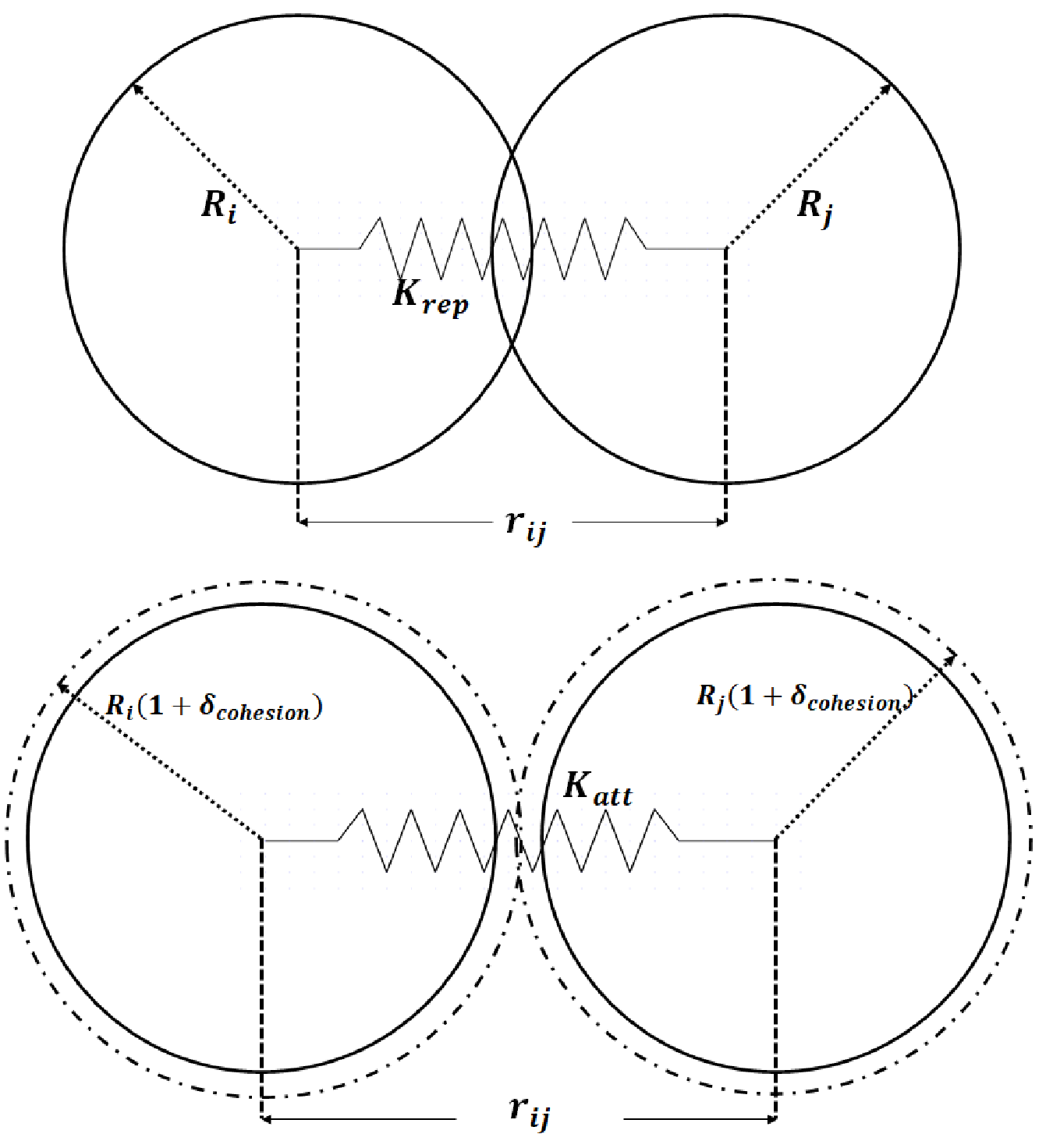}}
	\hspace{0.05\textwidth}
	\subfloat[Inter-particle forces between particles. \label{fig:force_plots}
	]{\raisebox{0.05\textwidth}{\includegraphics[width=.45\textwidth]{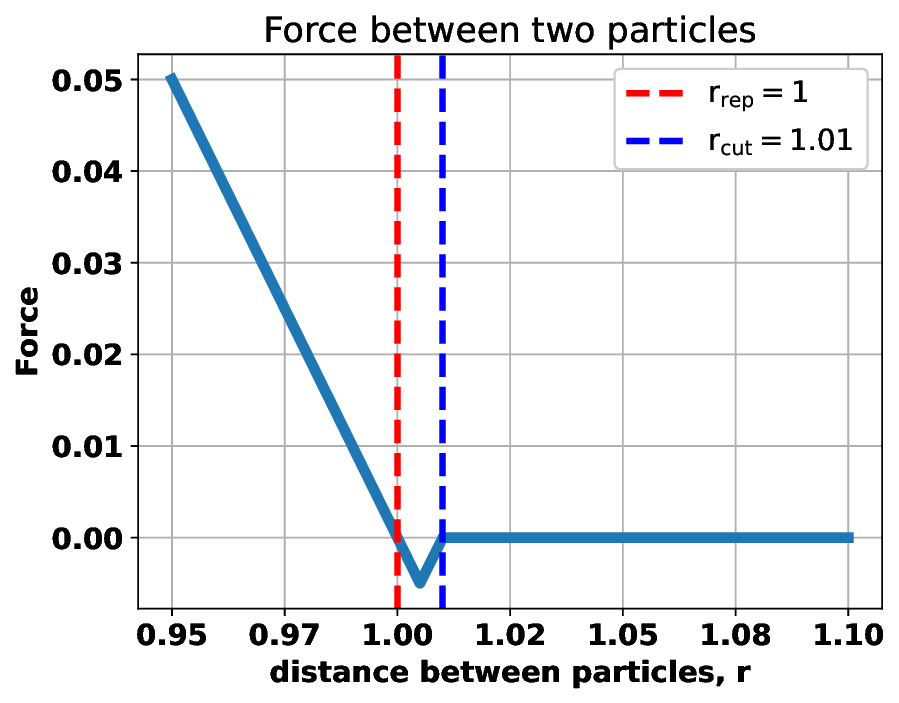}}}
        \caption{ 
		Figures illustrate the modified Hookean model that simulates the interaction between the particles in granular systems.
		In \cref{fig:illustrations}, the top figure illustrates the model used to simulate particle repulsion upon contact, while the bottom figure depicts the model used to simulate particle cohesion within the interaction range.
	In \cref{fig:force_plots}, the force on a particle during interaction with another particle is shown for $R_i=R_j=0.5$ and $K_{\text{rep}} = K_{\text{att}} = 1$. 
    The parameters are selected to ensure that the force function remains continuous. 
}
	\label{fig:spring}
\end{figure*}

The bi-disperse granular systems consist of two types of particles, $A$ and $B$, with radius $R_A$ and $R_B$.
The interaction between any two particles located at positions $\mathbf{r}_i$ and $\mathbf{r}_j$ is modeled using a modified Hookean model as shown in \cref{fig:spring} and \cref{eq:springmodel}.
As per the equation, the particles with a separation distance, $ r_{ij} \leq r_{\text{cut}}$,  will experience force only in the direction, $\mathbf{\hat{n}}_{ij}$, normal to the contact surface. 
The variable, $r_{\text{rep}}$, defines the maximum distance between the particles, $i$ and $j$, where they will experience repulsion.
Beyond the distance $r_{\text{rep}}$ and up to a distance of $ r_{\text{cut}}$, the particles will experience attractive force.  
The parameter $ r_{\text{cut}}$ is controlled using the cohesion length $\delta_{\text{cohesion}}$.
The length $\delta_{\text{cohesion}}$ describes the distance where the particles will experience attraction before coming into contact.
The parameters $K_{\text{rep}}$ and $K_{\text{att}}$ are the coefficients of repulsion and attraction, respectively, between two interacting particles $i$ and $j$ and are uniform across the system.

\begin{align}
	&\mathbf{F}_{ij}(\mathbf{r}_{ij}) =  
	\begin{cases}
		-K_{\text{rep}} (r_{ij}-r_{\text{rep}}) \mathbf{\hat{n}}_{ij}  & r_{ij} \leq r_{\text{rep}}\\
		-K_{\text{att}} (r_{ij}-r_{\text{rep}}) \mathbf{\hat{n}}_{ij} &  r_{\text{rep}} < r_{ij} \leq  .5r_{\text{cut}}\\
		K_{\text{att}} (r_{ij}-r_{\text{cut}}) \mathbf{\hat{n}}_{ij}  &  .5r_{\text{cut}} < r_{ij} \leq  r_{\text{cut}}\\
		0 & r_{\text{cut}} < r_{ij} 
	\end{cases} \label{eq:springmodel} \\
	\text{where} \qquad &\mathbf{r}_{ij} = \mathbf{r}_i-\mathbf{r}_j , \quad r_{ij} = |\mathbf{r}_{ij}|, \quad \text{and} \quad  \mathbf{r}_{ij} = r_{ij} \mathbf{\hat{n}}_{ij} \nonumber \\ 
	&r_{\text{rep}} = R_i + R_j , \quad r_{\text{cut}} = r_{\text{rep}} (1+\delta_{\text{cohesion}}) \quad  \text{and} \quad \forall i,j \in \{A,B\} \nonumber
\end{align}

The contact model also simulates the interaction between particles and rigid surfaces, such as boundary walls and indenters.
However, in these cases, the thickness of the rigid surface is set to zero, i.e., $R_i=0$, while $R_j$ remains non-zero for all $j \in \{A, B\}$.
Additionally, no cohesive forces are considered between the particles and rigid surfaces; therefore, $K_{\text{att}}=0$ for all interactions with rigid surfaces.

\subsection{Granular System Characterization}\label{sec:characterize}

Each granular system under examination is characterized by two scalar parameters that quantify the levels of disorder and cohesion.
The disorder in the systems arises from differences in the sizes of two-particle classes, as discussed in \cref{subsec:disorder}, thereby affecting the spatial arrangement and distribution of the particles.
These factors significantly impact the granular system response to external forces and their ability to form stable structures.

\subsubsection{Characterization of Disorder}\label{subsec:disorder}

The disorder of the granular system is quantified using a nonnegative scalar parameter, $D_{\text{size}}$, which governs the size of two-particle classes using \cref{eq:sizedisorder}.
The variable $\hat{d}$ represents the characteristic length of the simulation.
Effectively, The disorder parameter represents the size difference between the two types of particles, $A$ and $B$, in granular systems.
The greater the disorder, the larger the value of $D_{\text{size}}$, indicating a more significant variation in the particle sizes between the two types.

\begin{align}
	R_A = 0.5(1 + 0.1D_{\text{size}})\hat{d} \qquad \text{and} \qquad R_B = 0.5(1 - 0.1D_{\text{size}})\hat{d}  \label{eq:sizedisorder}
\end{align}

\subsubsection{Characterization of Cohesion}

The cohesion of the granular system is described by a scalar parameter $D_{\text{cohesion}}$, referred to as the cohesion factor in this work.
It is defined as the ratio between the attraction coefficient $K_{\text{att}}$ and the repulsion coefficient $K_{\text{rep}}$, as shown in  \cref{eq:cohesion}.
The cohesion factor is used to govern the balance between the strength of attractive and repulsive forces.

\begin{align}
	D_{\text{cohesion}} = \frac{K_{\text{att}}}{K_{\text{rep}}}  \label{eq:cohesion}
\end{align}

A higher value of $D_{\text{cohesion}}$ signifies a stronger attraction between particles once they are within the cutoff radius.
It is important to note that we keep the value of $D_{\text{cohesion}} < 1$ as values greater than $1$ would result in unrealistic behavior, where particles may bind too strongly, preventing proper movement or interaction and potentially leading to non-physical system behavior.

\subsection{Simulation Setup}\label{sec:simulate_setup}

All granular systems under study are modeled in a 2D simulation box, with the same number of particles and consistent fractions of particle classes $A$ and $B$, regardless of disorder or cohesion. 
The boundary condition is periodic on the sides and consists of a rigid wall at the bottom.
The simulations are conducted in a gravity-free environment to avoid the influence of particle weight-induced forces on particle interaction.
During the simulation, once the net forces on particles are computed using the contact model, a second-order differential equation, $\mathbf{F} = m\mathbf{\ddot r}$ is numerically integrated to update the particle velocities, $\mathbf{\dot{r}}$, and positions, $\mathbf{r}$.
Here, $\mathbf{F}$ is the net force and $m$ is the particle mass. 
This process is repeated until the system reaches equilibrium, where the net forces between particles are balanced, and the system stabilizes.
During the iterations, the number of particles and the size of the simulation box are kept constant.
The granular system setup and simulation of their response under uniaxial compression are performed using the open-source molecular dynamics package LAMMPS \cite{LAMMPS}. 
LAMMPS package uses the time integration algorithm based on the Stormer-Verlet symplectic integrator to solve the differential equation.

\subsubsection{Granular System Initialization}\label{sec:initialize}

\begin{figure*}
	\centering
	\subfloat[System with $D_{\text{size}} = 0.1$ \label{fig:disorder_01}]{\includegraphics[width=.35\textwidth]{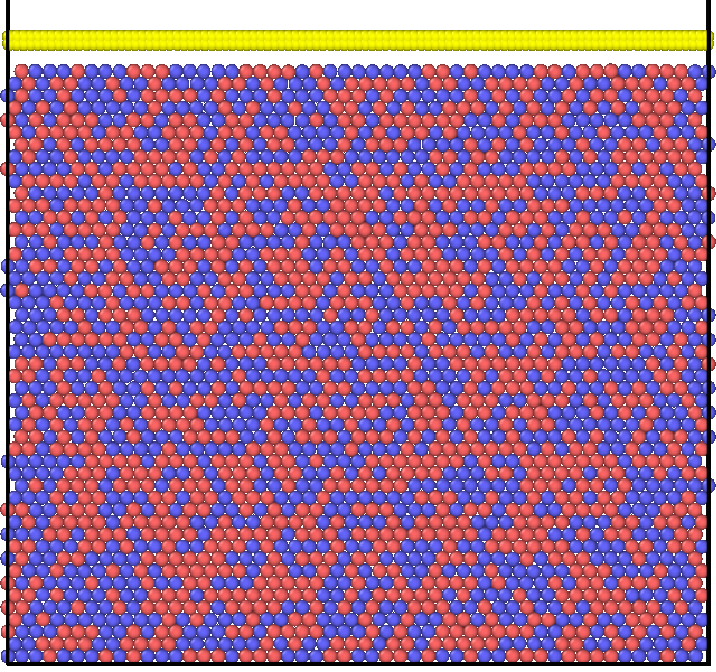}}
    \quad \quad 
	\subfloat[System with $D_{\text{size}} = 1$ 
	\label{fig:disorder_1}]{\includegraphics[width=.35\textwidth]{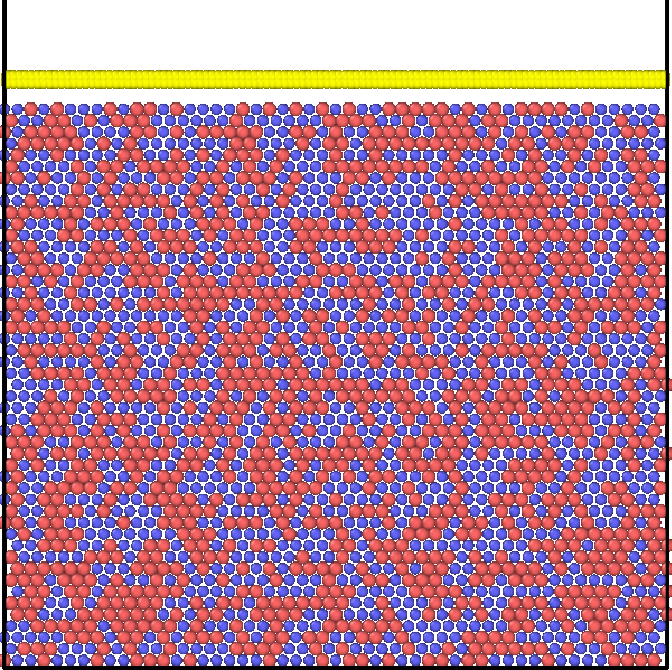}}
	\caption{
        The figure shows two cohesionless granular systems in their initial state, composed of particle classes $A$ (red) and $B$ (blue). 
        A planar indenter (yellow) is positioned above the particles, but initially, it does not apply force.
        Due to the significant size difference between the particles, the system in \cref{fig:disorder_1} is not jammed at the initial state.
        For Illustration, this figure shows systems containing approximately $2,500$ particles, although evaluations are conducted for larger systems with around $10,000$.
    }
	\label{fig:Initial}
\end{figure*}

The granular systems are initialized by placing the particles on hexagonal lattice points as shown in \cref{fig:Initial}.
The lattice constant $c$ is computed as $c=2/(\sqrt{3}r_{\text{max}}^2) $, where $r_{\text{max}} = \max{(R_A, R_B)}$. 
By systematically arranging the particles on a lattice using the computed lattice constant, overlap between particles is prevented.
This ensures the initial state represents an ordered system, which remains in equilibrium until external forces are applied.
However, it is essential to note that the particles in the systems at the initial state may not be tightly packed, and there might be some initial gaps between particles $A$ and $B$ (depending on the disorder parameter $D_\text{size}$), as shown in \cref{fig:disorder_1}.

Before applying external load, the simulation is run with the NVT (constant number of particles, volume, and temperature) ensemble for a few timesteps to drain kinetic energy and bring the systems to an equilibrium state.
The systems are at equilibrium if their kinetic energy is less than $10^{-9}$ units.

\begin{figure*}
	\centering
	\subfloat[System with $D_{\text{size}} = 0.1$ \label{fig:FC_no_positonal_disorder}]{\includegraphics[width=.45\textwidth]{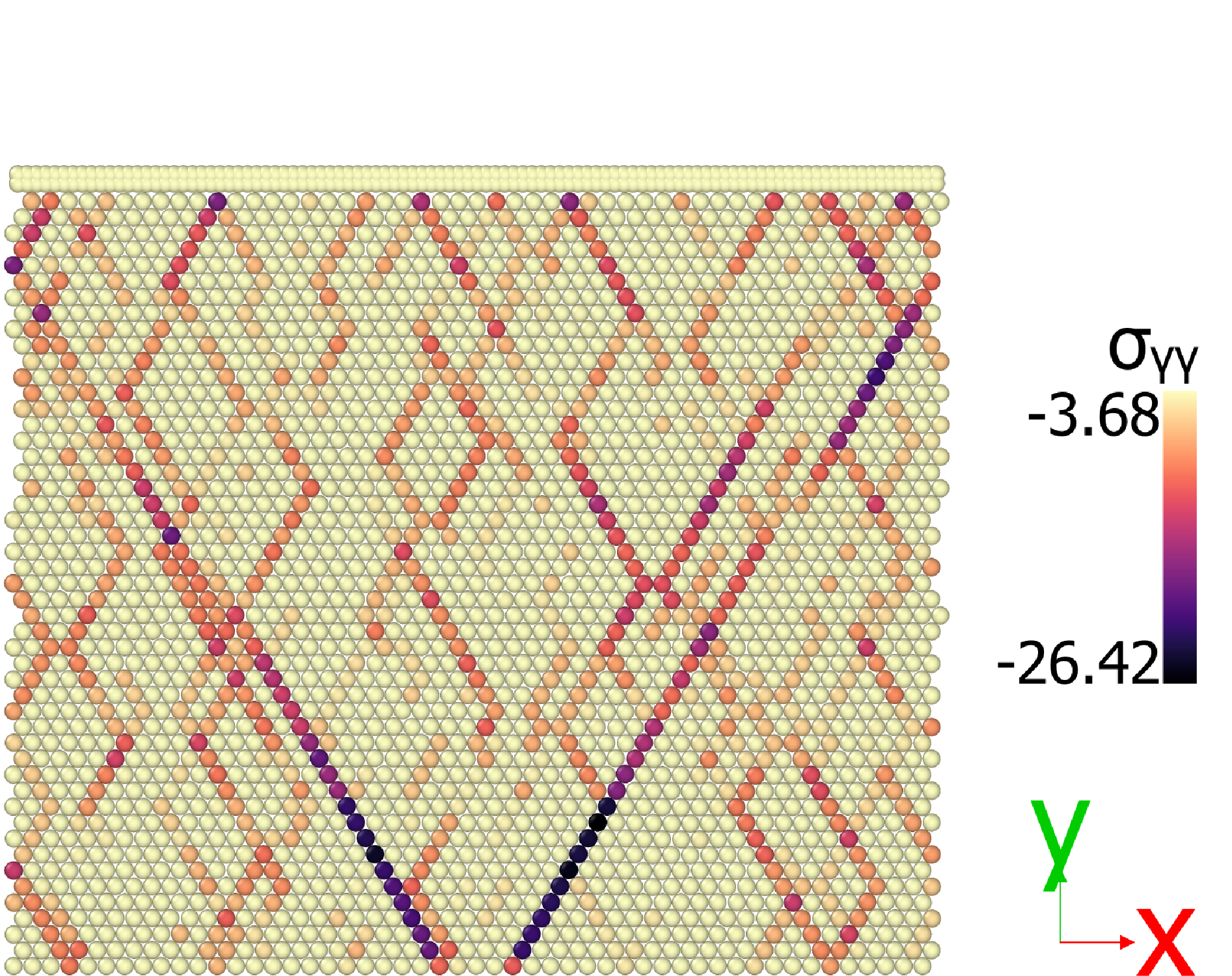}}
     \quad \quad
     \subfloat[System with $D_{\text{size}} = 1$ \label{fig:FC_positonal_disorder}]{\includegraphics[width=.45\textwidth]{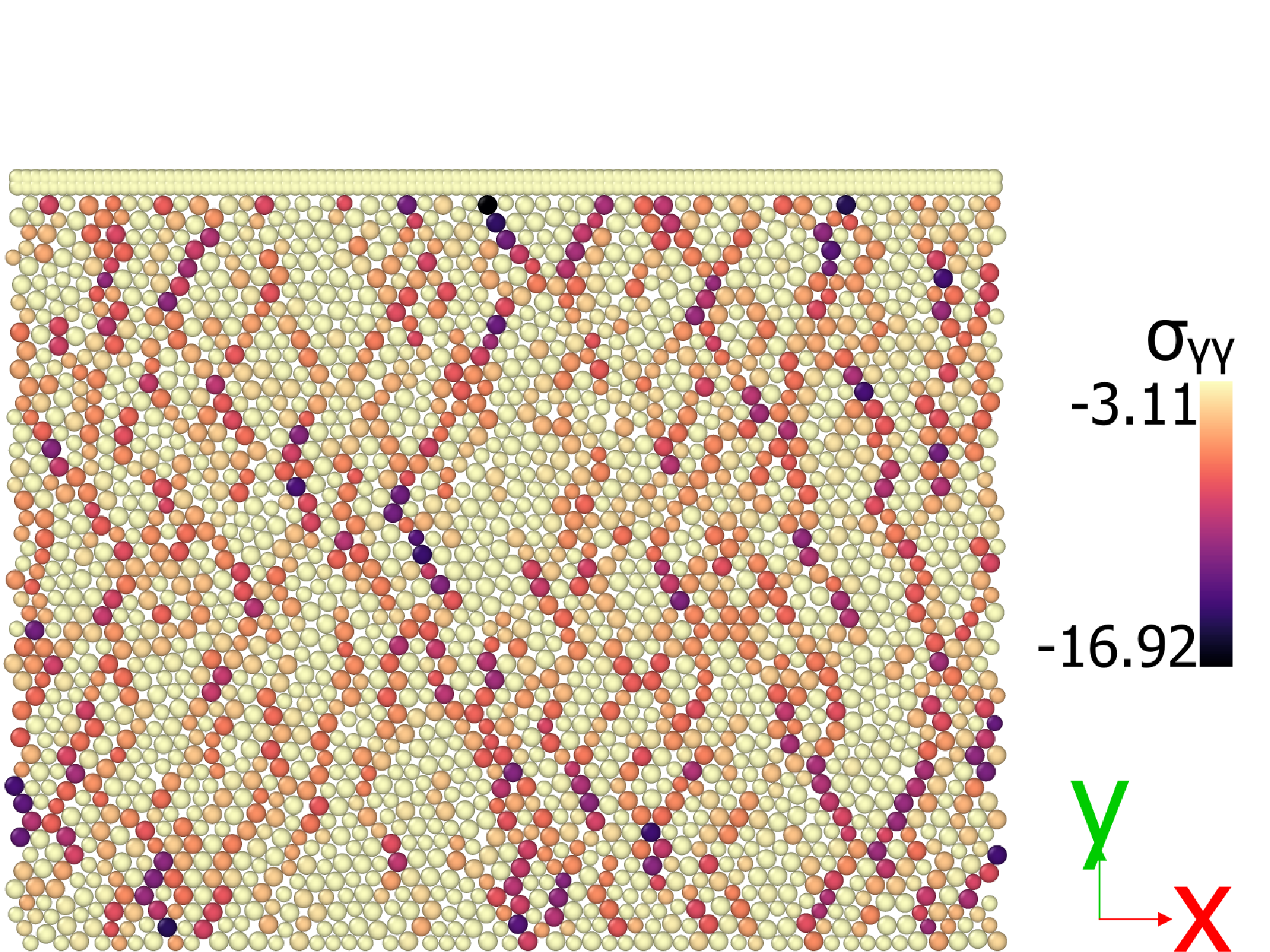}}
	\caption{
        The figure shows the stress \( \sigma_{yy} \) in cohesionless granular systems under uniaxial compression along the $y$ axis. 
        The stress values displayed are normalized unitless.
        The indenter applies a uniform pressure of $4$ units.
        Since LAMMPS outputs stress with tension-positive convention, the stress values are negative, representing particles in compression.
        }
	\label{fig:Final}
\end{figure*}

\subsubsection{Simulating Granular Response to Compression}\label{sec:simulate}

The systems are subjected to uniaxial compression using a rigid planar indenter, as shown in \cref{fig:Initial}, which applies a constant, uniform pressure on the top surface.
The indenter is designed using a collection of tiny particles, which form a rigid horizontal surface.
The indenter will keep compressing the systems until the particles exert enough force to resist the indenter load.
During the simulation, the Langevin thermostat, as described in \cite{schneider1978molecular}, is used to drain the kinetic energy out of the systems.  
Once the systems have reached equilibrium, particle stress tensor and force between particles are extracted and later used to characterize force chains. 

\subsection{Force Chains Characterizations}\label{sec:forcechain_criteria}

In the existing literature, multiple definitions have been proposed for the force chains using the stress and inter-particle forces obtained during simulations \cite{ardanza2013topological,peters2005characterization, huang2015community, nauer2020random, dijksman2018characterizing, krishnaraj2020coherent, bassett2015extraction}.
However, this study defines the force chains in three steps.
In the first step, particles with a compressive stress, $\sigma_{yy}$, exceeding the mean compressive stress of the system are identified. 
The identified particles are then treated as nodes of an undirected graph network.
The magnitude of the inter-particle forces is described as the weights of the edges. 
Describing force chains as graphs is helpful, as the characteristics of the chains can be analyzed using graph-based descriptors

\begin{figure*}
	\centering
	\subfloat[ Particles satisfying the mean stress criteria. 
	\label{fig:FC_preprocessed}]{\includegraphics[width=.4\textwidth]{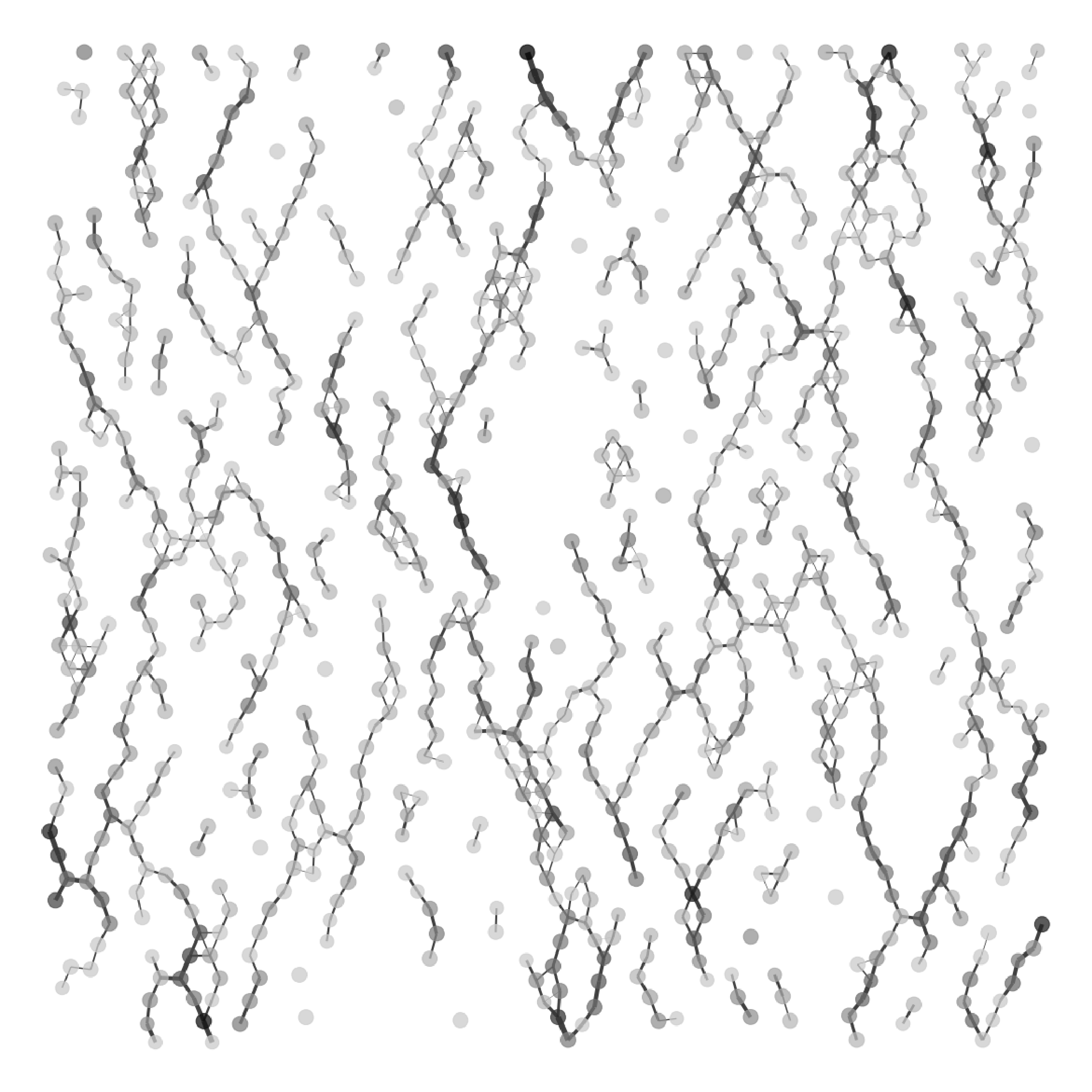}}
	\quad \quad
	\subfloat[ Particles after removing small subgraphs.
	\label{fig:FC_postporcessed}]{\includegraphics[width=.4\textwidth]{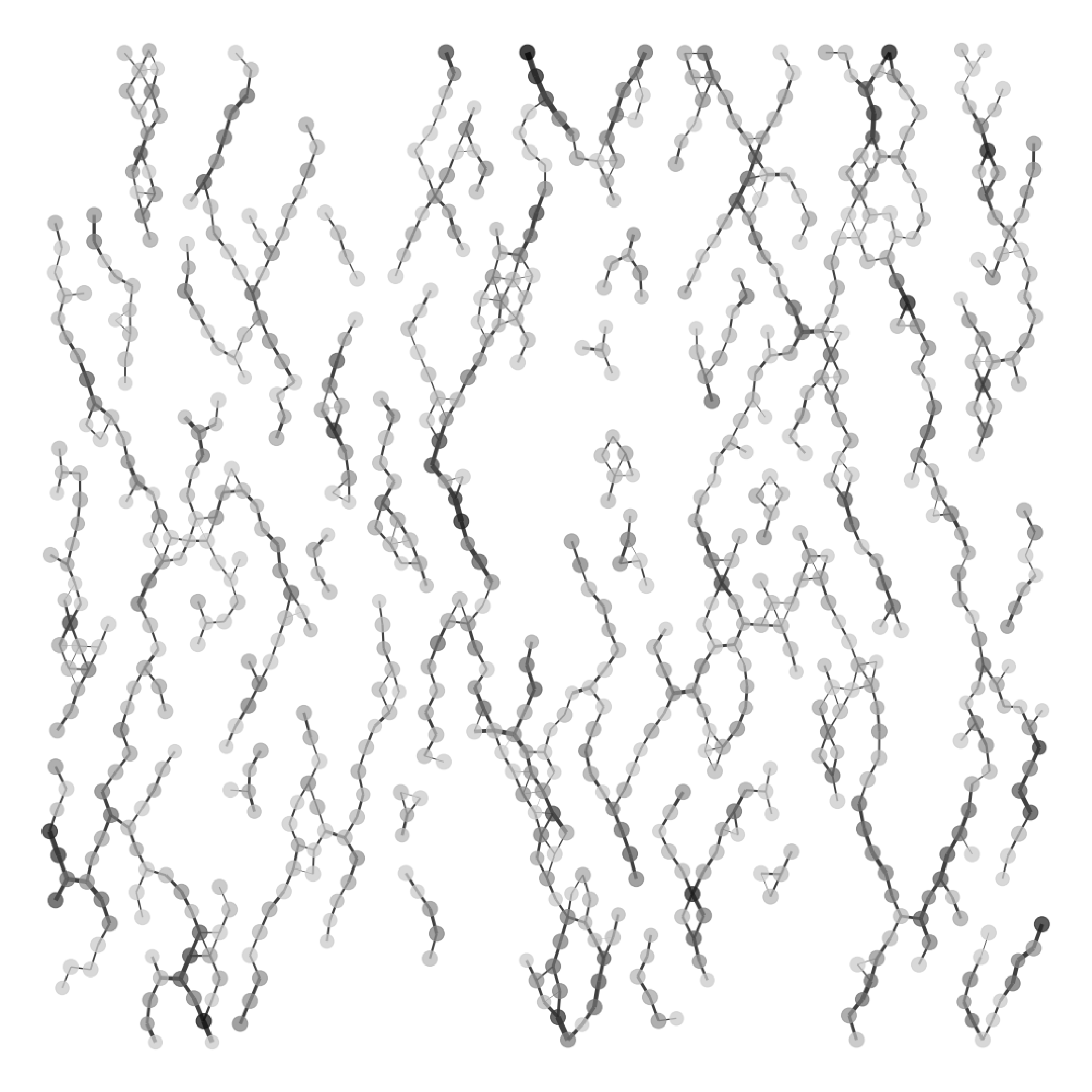}}
	\caption{
        The figure illustrates the force chains in a granular system with \( D_{\text{size}} = 1 \) and \( D_{\text{cohesion}} = 0 \) (see \cref{fig:FC_positonal_disorder}). 
        The force chains are represented as a graph, with particles as nodes and inter-particle forces as edges. 
        Edge thickness indicates force magnitude, while node color represents scaled stress, with black nodes having the highest stress. 
        The \cref{fig:FC_preprocessed} shows a network with particles meeting the mean stress and inter-particle force criteria, containing several small subgraphs. 
        After removing the small subgraphs, the resulting network, shown in \cref{fig:FC_postporcessed}, represents the force chains in the granular system.
		}
	\label{fig:force-chains}
\end{figure*}

In the next step, an edge is removed if the corresponding inter-particle force is below the average inter-particle force.
This process filters out the weaker forces, which mainly provide lateral support to the chains rather than contribute to the transfer of the applied load.
This process results in a more distinct force chain network and ensures that adjacent force chains, which are not significantly interacting, are not incorrectly counted as a single chain. 
For the granular system in \cref{fig:FC_positonal_disorder}, the corresponding force chain network after the first two processing steps is shown in \cref{fig:FC_preprocessed}.

The resulting graph network, representing force chains, may consist of multiple subgraphs, each representing an independent chain as seen in \cref{fig:FC_preprocessed}.
However, some subgraphs contained very few nodes relative to the system size. 
These small subgraphs, which arise from stress concentrations, do not behave like force chains, as force chains typically span multiple-grain scales and form long, continuous structures.
Hence, the subgraphs with less than four nodes were removed from the graph.
For the granular system in \cref{fig:FC_positonal_disorder}, the final force chain network is shown in \cref{fig:FC_postporcessed}.

\section{Numerical Experiments}\label{sec:Ex_setup}

Multiple bi-disperse granular systems have been set up with the boundary conditions and external loading as detailed in  \cref{sec:simulate_setup}.
The systems consist of $10,000$ discs initially placed on hexagonal lattice points.
The lattice constant $c$ depends on the value of the larger radius of the particles, as discussed in  \cref{sec:initialize}.

For the current simulation, all particles are treated as having unit masses.
The value of the characteristic length of the simulation, $\hat{d}$, is set as $1$.
All other variables, such as radius, distance, and simulation box size, are multiples of $\hat{d}$. 
The value of $K_{\text{rep}}$ is set as $1\times10^5$ for all systems, and the values of pressure and force are discussed in multiple of $K_{\text{rep}}$ units.
The indenter applies a uniform pressure of value $4\times10^{-5} K_{\text{rep}}$ on the systems.
The timestep for the system is set as $0.0001$ units, with the characteristic time scale as $\sqrt{1/K_{\text{rep}}}$.

The effects of disorder and cohesion on the force chains within the systems are investigated separately. 
The influence of disorder is examined by studying multiple cohesionless granular systems with disorder parameter, $D_{\text{size}}$, within the range of $[0,1]$.
The effects of cohesion on the force chains are studied by varying the cohesion factor, $D_{\text{cohesion}}$ within the range $[0,1]$ and for the systems with disorder values of $D_{\text{size}} = \{0.05, 0.1, 1\}$.

\section{Observations}\label{sec:results}

Various graph descriptors enable the evaluation of multiple properties of force chains; however, it remains challenging to interpret how these properties relate to the physical phenomena observed in the system.
In this study, we focus on the fraction of particles, $N_F$, that are part of the force chains as a metric for assessing the presence of these chains.
As per the criteria discussed in  \cref{sec:forcechain_criteria}, the particles in the force chains exist exclusively in a well-defined network that spans multiple particle scales, transmitting the external load.
Hence, the metric $N_F$ can be used to quantify whether the force chains are present in the systems.
The $N_F$ value of a granular system with fewer force chains will be smaller than that of a system with more force chains.
If $N_F=0$, then no force chains exist in the system.
This work uses the metric $N_F$ to examine how disorder and cohesion influence the formation of force chains in granular systems.
Furthermore, we also assess the conditions under which variations in particle size result in size disorder versus when they lead to positional disorder.

\subsection{Effect of Disorder}

\begin{figure*}
	\centering
	\includegraphics[width=.5\textwidth]{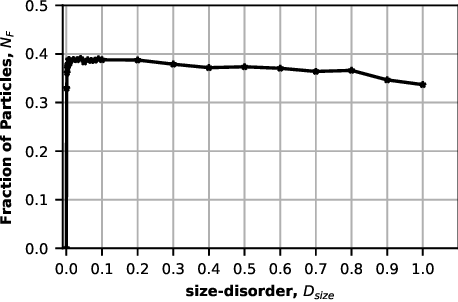}
	\caption{
		The figure shows the effect of disorder on the force chains in the granular systems.
		As the disorder increases, the force chains appear in the granular systems.
		For the range, $D_{\text{size}} = [0,0.1]$, the number of particles in force chains increases suddenly and then remains constant for a while.
		Beyond the range, $D_{\text{size}} = 0.1$, the number of particles gradually decreases as the positional disorder becomes significant. 
	}
	\label{fig:disorder-effects}
\end{figure*}

\begin{figure*}
    \centering
    \subfloat[For $D_{\text{size}}=0$ \label{fig:RDF0}]{\includegraphics[width=.45\textwidth]{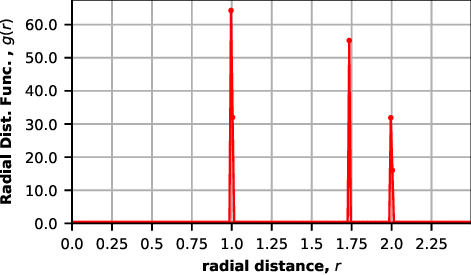}}
    \quad \quad
    \subfloat[For $D_{\text{size}}=0.005$\label{fig:RDF005}]{\includegraphics[width=.45\textwidth]{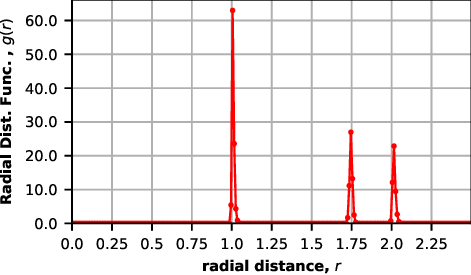}} 
    \\
    \subfloat[For $D_{\text{size}}=0.01$  \label{fig:RDF01}]{\includegraphics[width=.45\textwidth]{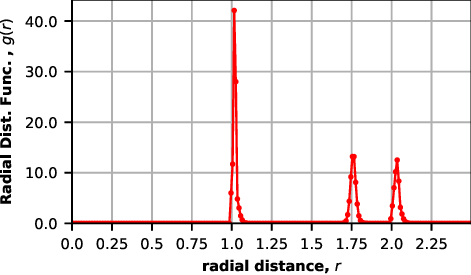}}
    \quad \quad
    \subfloat[For $D_{\text{size}}=0.1$ \label{fig:RDF1}]{\includegraphics[width=.45\textwidth]{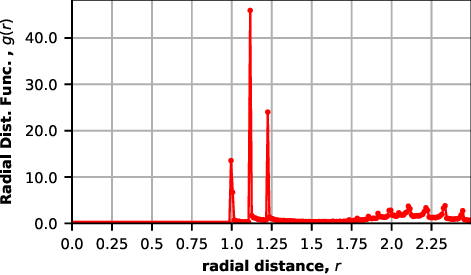}}
    
    \caption{
		The figure shows the radial distribution functions (RDF) plots for granular systems with different disorders.
		Here, the radial distribution functions are used to characterize the positional disorder of the systems.
	\cref{fig:RDF0} also represents the RDF of a hexagonally packed system. 
	\cref{fig:RDF005}, \cref{fig:RDF01} suggests that the positional disorder in the systems with a disorder of  $D_{\text{size}}=0.05$ and $D_{\text{size}}=0.1$ are not significant.
	\cref{fig:RDF1}, suggests significant positional disorder in the system with $D_{\text{size}}=1$.
			}
	\label{fig:positional-disorder}
\end{figure*}

The effects of the disorders on the force chains are evaluated by comparing the value of $N_F$ against the value of the disorder parameter $D_{\text{size}}$, see \cref{fig:disorder-effects}. 
As the disparity in particle sizes was increased to enhance the disorder, it ultimately affected the packing, resulting in a different type of disorder.
It was observed that different types of disorders had distinct effects on the characteristics of force chains.
To classify the type of disorder present in a granular system, we analyzed the particle packing using radial distribution functions (RDF) $g(r)$.
The RDF describes the spatial distribution of particles in a system, mainly how particle density varies as a function of distance from a reference particle. 
For a perfectly ordered hexagonal lattice system, where no disorder exists, the RDF is shown in \cref{fig:RDF0}.

As shown in \cref{fig:disorder-effects}, for $D_{\text{size}}=0$, $N_F$ is $0$, indicating that there are no force chains in the granular system. 
The absence of force chains in systems with $D_{\text{size}}=0$ is due to the perfectly ordered system, where particles are arranged in a hexagonal lattice, and all particles have the same properties.
The external load from uniaxial compression is uniformly distributed across the granular system, preventing the formation of force chains in such an ordered packing.

Later, with a slight increase in $D_{\text{size}}$, a sharp rise in $N_F$ indicates that force chains will be observed in systems even with minor disorders. 
Subsequently, the metric $N_F$ remains constant till $D_{\text{size}}=0.1$, indicating that the number of particles involved in force chains stays nearly unchanged despite variations in other characteristics of force chains.
For $D_{size} \leq 0.1$, the variation in size mainly leads to size disorder with minimal positional disorder.
The disorder type has been verified using RDF plots as shown in \cref{fig:RDF005} and \cref{fig:RDF01} where the packing closely approximates a hexagonal lattice packing, suggesting that the system is near an ordered state.

Beyond $D_{\text{size}}=0.1$, the value of $N_F$ starts to decrease, suggesting fewer particles participate in forming force chains in the systems.
Further visualization shows that granular systems with $D_{\text{size}}>0.1$ exhibit shorter and more fragmented force chains.
These fragmented force chains are caused by the large disparity in particle sizes, which induces significant positional disorder, as seen in \cref{fig:RDF1}.
The RDF plot in \cref{fig:RDF1} differs significantly from the other plots, indicating that the particle packing is no longer close to an ordered state.

Figures \cref{fig:disorder-effects} and \cref{fig:positional-disorder} illustrate that the type of disorder influences force chains significantly. 
Size disorder leads to more unified and continuous chains, while positional disorder causes the chains to become more fragmented.

\subsection{Effect of Cohesion}

\begin{figure*}
	\centering
	\includegraphics[width=.45\textwidth]{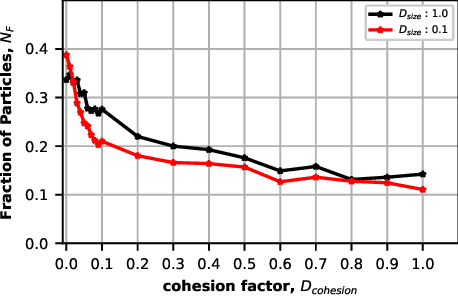}
	\caption{
		The figure shows the effect of cohesion on the force chains in the granular systems with two different disorders.
		The value of disorder parameter $D_{\text{size}}=0.1$ represents the system with a minor disorder and no positional disorder.
		The value of disorder parameter $D_{\text{size}}=1$ represents the systems with significant positional disorder.
		The plots suggest that increased cohesion causes force chains to disappear in both positional ordered and disordered systems. 
	}
	\label{fig:cohesion-effects}
\end{figure*}

The effect of cohesion on the force chains is evaluated by comparing the values of $N_F$ against the values of the cohesion factor $D_{\text{cohesion}}$. 
As shown in \cref{fig:cohesion-effects}, the value of $N_F$ decreases with the increase in the cohesion factor, irrespective of the type of disorder present in granular systems.
The plot suggests that the force chains will disappear in the systems as cohesion between the particles is increased.

\begin{figure*}
    \centering
    \subfloat[Systems with no significant positional disorder, $D_{\text{size}}=0.1$ \label{fig:cohesion_stress1}]{\includegraphics[width=.45\textwidth]{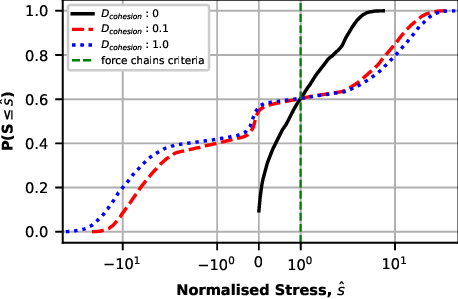}}
    \quad
    \subfloat[Systems with significant positional disorder, $D_{\text{size}}=1$ \label{fig:cohesion_stress2}]{\includegraphics[width=.45\textwidth]{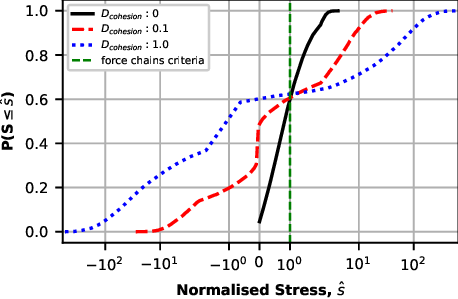}}

    \caption{ The cumulative distribution plots of the normalized stress, $\sigma_{yy}$, in granular systems are presented for cases with minimal positional disorder, \cref{fig:cohesion_stress1} and significant positional disorder, \cref{fig:cohesion_stress2}. Particles with normalized stress values greater than $1$ are identified as part of force chains unless they belong to a very small subgraph. Negative normalized stress values indicate particles under tension due to cohesion.
    As the cohesion strength increases within the system, more particles contribute to load re-distribution. 
    The load re-distribution reduces the number of particles directly involved in significantly large load transfer compared to nearby particles, leading to a decrease in the formation of force chains, as illustrated in \cref{fig:cohesion-effects}.
    }
    \label{fig:cohesion_distribution}
\end{figure*}

To further investigate the causes of force chain disappearance, we analyze the cumulative distribution plots of normalized stress in the granular systems.
In \cref{fig:cohesion_stress1} and \cref{fig:cohesion_stress2}, according to our force chain criteria, particles with normalized stress values less than one are not part of force chains.
The cumulative distribution plots show that as cohesion increases in a granular system, more particles exhibit negative normalized stress values, indicating that these particles are under tension.
Therefore, fewer particles satisfy the defined threshold criteria for force chains, resulting in fewer force chains. 
A reduction in force chains with increased cohesion was also observed by \cite{wang2017force}. 
They noted that cohesion increased the presence of weak forces within the system.
Cohesive bonds enhance particle interactions even among particles not in direct contact, redistributing the applied load and increasing the prevalence of weak force interactions.

In cohesionless systems, there are no cohesive bonds to create interactions between particles that are not in direct contact.
This results in the applied load primarily transmitting through distinct and localized force chains, with minimal re-distribution across weaker contacts.
Hence, Cohesionless systems exhibit more pronounced and concentrated force chains.

\section{Concluding Remarks}\label{sec:conclusion}

This work discussed the independent role of particle packing, size variation, and cohesion on the force chains in bi-disperse granular systems under uniaxial compression.
The particles in the granular systems are initialized on hexagonal lattice points with zero initial velocities, and then an indenter is used to apply uniaxial compression.
The interaction between the particles is simulated using a modified Hookean model.
The granular packings were characterized by a cohesion factor and a disorder factor.
The disorder factor was quantified by the difference in the size of the two classes of particles present in bi-disperse granular systems.

Multiple granular systems with different disorders and cohesion were simulated.
Once the systems under uniaxial compression reach equilibrium, the stresses, in the direction of uniaxial compression, and forces between the particles are used to observe the force chains.
Once the force chains are extracted, the number of particles in the force chains is calculated. 
The fraction of particles in the granular system present in the force chains is used as a metric to characterize the presence of force chains in the granular system.
Both disorder and cohesion significantly affected the presence of force chains in granular systems. 

As the disparity in particle sizes increased, two distinct types of disorder were observed, each leading to different characteristics of force chains.
The first type of disorder, size disorder, occurred when the disorder was primarily due to differences in particle sizes while the packing remained nearly ordered. 
It was observed that force chains can form in systems even with small size disparities while they remain almost ordered.
The second type of observed disorder, positional disorder, arose when significant variations in particle size led to irregular packing.
The presence of positional disorder resulted in more fragmented force chains.
A radial distribution function was used to identify the type of disorder.

Further, it was observed that the force chains start to disappear with an increase in cohesion, regardless of the type of disorder. 
In cohesionless, disordered granular systems, particles that are close to one another yet not in direct contact do not interact. 
Conversely, particles within the interaction range but not in contact in cohesive granular systems begin to interact significantly. 
The increased particle interactions due to cohesion redistributed the applied load, resulting in more weak forces and fewer force chains.

The proposed approach to characterizing disorder and cohesion in this work allowed for an investigation of the effects of particle size and cohesion on force chains independent of the packing in granular media.
The characterization method and computational setup can be extended to explore the independent impacts of friction, particle shape \cite{macedo2023shape}, and various external loading conditions. 
More advanced graph-based descriptors can be employed to analyze force chain characteristics. By leveraging machine learning techniques, these descriptors could provide more accurate predictions of the material bulk-scale response.
More complex data science models, such as principal component analysis and diffusion maps, can be employed to study the evolution of the force chains similar to microstructure evolution discussed in \cite{desai2024trade}.

The knowledge of relationships between micro-scale properties and force chains can help build data science methods to predict localized high stresses in granular media, following  similar ideas in polycrystals \citep{shrivastava2022predicting}. 
Further, the knowledge of the role of micro-scale properties on force chain characteristics and, eventually, macro-scale response can help design granular media with desired characteristics similar to the models proposed in \cite{shrivastava2024bayesian}.

%%%%%%%%%%%%%%%%%%%%%
%%%%%%%%%%%%%%%%%%%%%
%%%%%%%%%%%%%%%%%%%%%
%%%%%%%%%%%%%%%%%%%%%
\begin{acknowledgments}
    We acknowledge financial support from ARO (MURI W911NF-19-1-0245, MURI W911NF-24-2-0184) and NSF (DMREF 2118945, RTG 2342349), and NSF for XSEDE computing resources provided by Pittsburgh Supercomputing Center.
\end{acknowledgments}

% References
\newcommand{\etalchar}[1]{$^{#1}$}

\end{document}